\documentclass[twocolumn,10pt]{asme2e}

\usepackage{graphicx}
\graphicspath{ {./images/} }
\usepackage{amsmath}
\usepackage{mathtools}
\usepackage{xcolor}
\usepackage{tabularx}
\newcolumntype{s}{>{\hsize=.3\hsize}X}
\usepackage{booktabs}
\usepackage{caption}
\usepackage{subcaption}
\usepackage[symbol]{footmisc}
\usepackage{threeparttable}

\confshortname{IDETC/CIE 2022}
\conffullname{the ASME 2022 International Design Engineering Technical Conferences \&\\
              Computers and Information in Engineering Conference}

\confdate{August 14-August 17}
\confyear{2022}
\confcity{St. Louis}
\confcountry{USA}

\papernum{DETC2022/MESA-90132}

\title{CONCEPTUAL DESIGN OF CELLULAR AUXETIC SYSTEMS WITH PASSIVE ADAPTATION TO LOADING} 
\author{Joshua Prendergast\thanks{M.S. Student, Department of Mechanical and Aerospace Engineering. ASME Student Member}
    \affiliation{
	University at Buffalo\\
	Buffalo, New York 14260\\
    Email: jmprende@buffalo.edu
    }	
}

\author{Manaswin Oddiraju\thanks{Ph.D. Student, Department of Mechanical and Aerospace Engineering. ASME Student Member}
    \affiliation{
	University at Buffalo\\
	Buffalo, New York 14260\\
    Email: moddiraj@buffalo.edu
    }	
}

\author{Mostafa Nouh\thanks{Associate Professor, Department of Mechanical and Aerospace Engineering. ASME Professional Member}
    \affiliation{
	University at Buffalo\\
	Buffalo, New York 14260\\
    Email: mnouh@buffalo.edu
    }	
}

\author{Souma Chowdhury\thanks{Associate Professor, Department of Mechanical and Aerospace Engineering. ASME Professional Member. Corresponding Author.}
    \affiliation{
	University at Buffalo\\
	Buffalo, New York, 14260\\
    Email: soumacho@buffalo.edu
    }	
}

\begin{document}
\maketitle
\begin{abstract}
    Auxetics refer to a class of engineered structures which exhibit an overall negative Poisson's ratio. These structures open up various potential opportunities in impact resistance, high energy absorption, and flexible robotics, among others. Interestingly, auxetic structures could also be tailored to provide passive adaptation to changes in environmental stimuli -- an adaptation of this concept is explored in this paper in the context of designing a novel load-adaptive gripper system. 
    Defining the design in terms of repeating parametric unit cells from which the finite structure can be synthesized presents an attractive computationally-efficient approach to designing auxetic structures. This approach also decouples the optimization cost and the size of the overall structure, and avoids the pitfalls of system-scale design e.g., via topology optimization. In this paper, a surrogate-based design optimization framework is presented to implement the concept of passively load-adaptive structures (of given outer shape) synthesized from auxetic unit cells. Open-source meshing, FEA and Bayesian Optimization tools are integrated to develop this computational framework, enhancing it adopt-ability and extensibility. Demonstration of the concept and the underlying framework is performed by designing a simplified robotic gripper, with the objective to maximize the ratio of towards-load (gripping) horizontal displacement to the load-affected vertical displacement. Optimal auxetic cell-based design generated thereof is found to be four times better in terms of exhibited contact reaction force when compared to a design obtained with topology optimization that is subjected to the same specified maximum loading. 
\end{abstract}

\begin{nomenclature}
\entry{$\delta$}{Unit cell diagonal rib flat thickness}
\entry{$\theta_{\textrm{fixed}}$}{Angle spanning fixed edge on arch of gripper from circular arch center}
\entry{$\theta$}{Unit cell rib angle with respect to vertical}
\entry{$\nu$}{Poisson's ratio}
\entry{$\sigma_y$}{Yield strength of gripper material}\entry{$\rho_g$}{Gripper material density}
\entry{$a$}{Vertical distance from base of gripper jaws to ellipse center}
\entry{$a_1,a_2,b_1,b_2$}{Number of unit cells used to mesh the $y_1,y_2,x_1,x_2$ dimensions of the gripper}
\entry{$A_p$}{Payload cross sectional area}
\entry{$b$}{Horizontal thickness of gripper jaws}
\entry{$d$}{Depth of gripper, cells, and payload in the out of plane direction}
\entry{$E$}{Young's Modulus of gripper and payload}
\entry{$F_{\textrm{in}}$}{Gripper input actuation force at top surface}
\entry{$F_\textrm{out}$}{Gripper reaction force at payload contact}
\entry{FEM}{Finite Element Model}
\entry{FS}{Factor of safety}
\entry{$f$}{Design optimization objective function}
\entry{$g_i$}{$ i^{\rm{th}} $ constraint of design optimization problem}
\entry{$h_c$}{Unit cell half base width}
\entry{$\ell$}{Unit cell inclined rib length}
\entry{$L_p$}{Payload rectangular length}
\entry{NPR}{Negative Poisson's ratio}
\entry{$r$}{Gripper arch radius}
\entry{$t_g$}{Gripper thickness}
\entry{$t_c$}{Unit cell member thickness}
\entry{$\vec{u}$}{Displacement vector with x-component $u$ and y-component $v$}
\entry{$v_{\textrm{induced}}$}{Induced displacement at top surface of gripper}
\entry{$W_g$}{Overall width of gripper}
\entry{$w_p$}{Payload width}
\entry{$\vec{x}$}{Vector of design variables}
\end{nomenclature}

\section{Introduction}
\subsection{Overview}
Metamaterials are a class of synthetically designed (typically but not necessarily periodic) structures which produce unique mechanical, acoustical, magnetic, or thermal properties which are not easily found in nature. Auxetic metamaterials are designed to have a \textit{negative Poisson's ratio} (NPR) and therefore expand in the lateral direction when subject to a tensile load \cite{AuxeticMetamaterialsAndStructures}. Auxetic metamaterials can provide advantages over traditional materials when compared on the basis of indentation resistance, fracture toughness, vibration transmission, and more \cite{ThreeDecades}. They also have practical applications in the design of compliant mechanisms, soft robotics, filtration systems, and cushions \cite{ThreeDecades}.

Auxetic metamaterials can be broadly classified into four groups, namely the re-entrant type \cite{ReEntrant2D}, the chiral type \cite{ChiralReview}, the rotating polygon type and the planar sheet type. Among these, re-entrant auxetic structures are more popular owing to their simple geometry. Auxetics have prominent advantages over traditional materials when used in applications that benefit from indentation resistance, high fracture toughness, and wave attenuation. The indentation resistance of an isotropic material is proportional to $E/(1-\nu^2)$ when a uniform indentation pressure is assumed \cite{Timoshenko}. In the limit as $\nu \to -1$, indentation hardness tends to go to infinity for the case of classical isotropic elasticity. Auxetic materials derive their indentation and impact resistance by virtue of their NPR, since material flows into the contact region as a result of lateral contraction, which is called densification \cite{LateralThinking}. Most auxetic metamaterials are anisotropic \cite{liu2019mechanical}, so the isotropic relationship does not explicitly hold in all cases. Similarly, re-entrant auxetic foams have been compared to their conventional counterparts and have been shown analytically and experimentally to improve fracture toughness \cite{FractureToughness}. The anisotropic behavior of a repeating cellular structure also impacts the transmission of waves throughout a system, and resulting band gaps are determined by the unit cell parameters \cite{billon2017mechanics}. More generally, it has also been shown that auxetic foams have better acoustic absorption than conventional foams \cite{ThreeDecades}.

\begin{figure*}[t]
\includegraphics[width= \linewidth]{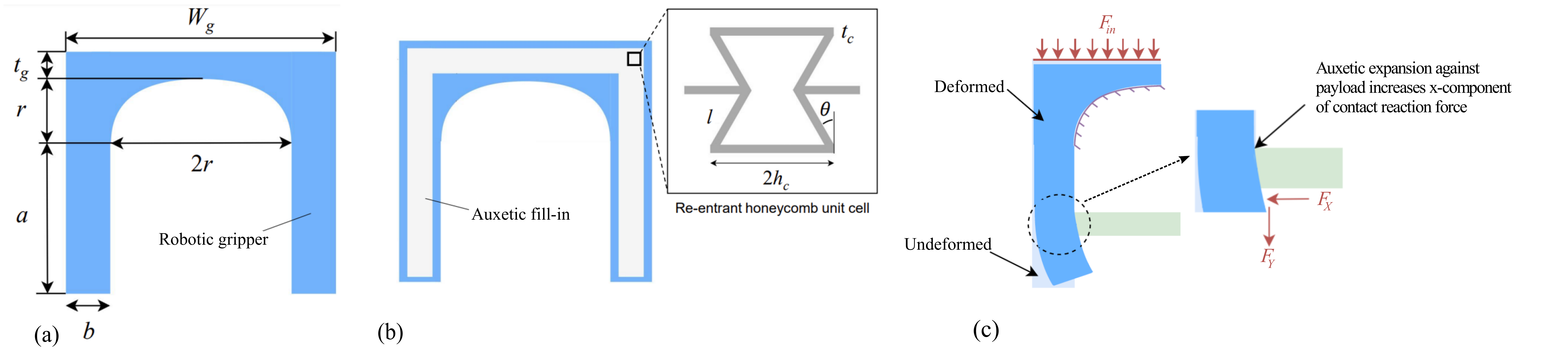}
\centering
\confdate{}\caption{Geometric parameters corresponding to (a) the robotic gripper and (b) the 2D re-entrant honeycomb unit cell and (c) the undeformed and deformed auxetic gripper}
\label{fig:Gripper_Cell}
\end{figure*}
Auxetic metamaterials have been used in the design of protective sports equipment and filtration systems, and have inspired the design of displacement amplifying compliant mechanisms \cite{duncan2018review}. The improved response of auxetic materials subject to impact loading when compared to traditional materials makes the use of auxetics in helmets and pads advantageous, in addition to the design benefiting from being lightweight \cite{Sports}. Similarly, auxetics may provide advantages in the manufacturing of body armor \cite{LateralThinking}, and have been experimentally and numerically investigated in response to blast loading motivated by protecting military vehicles from improvised explosive devices \cite{ImpactAndBlast}. Furthermore, porous auxetic materials have been used to fabricate active filtration devices whose size can controlled by the applied strain \cite{filter}. Compliant mechanisms which exhibit auxetic properties have also been developed and can be used as displacement amplifiers which offer alternative approaches to the design of auxetic systems when compared with the repeating unit cell concept \cite{MechanismAuxetic}. A detailed survey of auxetic applications is provided in references \cite{ThreeDecades} and \cite{AuxeticMetamaterialsAndStructures}.

Traditionally, the design of metamaterials with a target functionality has relied on heuristics and designer intuition \cite{FlexibleMechanicalMetamaterials, TopOp_CAD_2017} with few rigorous approaches existing that derive shapes which are expected to produce NPR either at a unit cell or system level. Reliance on only designer input is time consuming, severely restricts the likely design space, and generally results in sub-optimal designs. Moreover, the types of mechanisms that can apply auxetic metamaterials may be limited if the designer has to manually mesh the geometry with unit cells. Some work has been done in the development of automated approaches to the design of auxetic systems. For example, auxetic behavior is not limited to regularly repeating unit cell structures, and the design concept has been generalized to use optimization algorithms in the development of auxetic systems based on disordered networks \cite{DisorderedNetworks}. This approach can be less computationally expensive than full topology optimization for simple cases but for scenarios where a large nodal network is used, this method is likely to be as expensive as topology optimization since the optimization operates directly on the nodes.

Optimization of complex auxetic structures usually relies on topology optimization which is a computationally expensive approach and is typically focused on simple, albeit novel systems, not generalized shapes and designs \cite{TopOp_ArcApplMech_2015, TopOp_CAD_2017, CompliantMicromechanisms}. The limits of topology optimization are well explored. They include high computational cost and rough resultant shapes which usually require manual post processing \cite{sigmund2013topology} . Limited investigation into the use of auxetic metamaterials in the design of more complex systems such as manufacturable soft robotic grippers has been conducted \cite{SoftAuxeticGripper}. However, no general design framework was proposed or implemented in this work and the results only demonstrate a proof of concept. To our knowledge, a generalized framework for the automated design of practical auxetic systems which benefit from passive adaptation to loading has not yet been presented to the engineering design research community. Moreover, no generalized parametric optimal design schemes for auxetic systems have been published which robustly address the issue of computational cost. The goal of this paper is to build on previous research to introduce a robust optimal design framework that can be used for the automated development of auxetic systems.

\subsection{Auxetic structure with passive loading adaptation}

\begin{figure}[]
\begin{subfigure}{\linewidth}
\includegraphics[width= \linewidth]{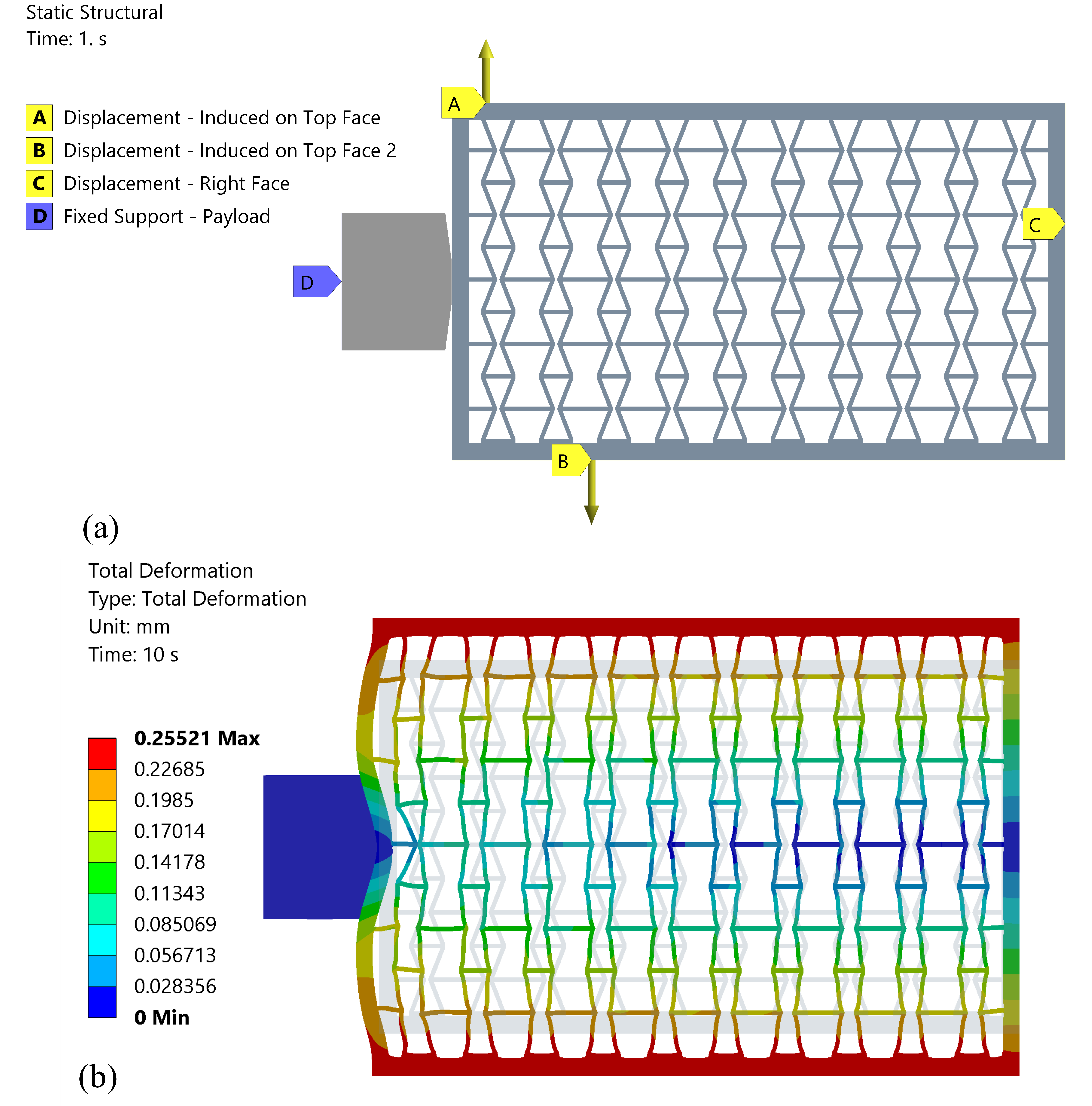}
\centering
\end{subfigure} \\
\begin{subfigure}{\linewidth}
\includegraphics[width= \linewidth]{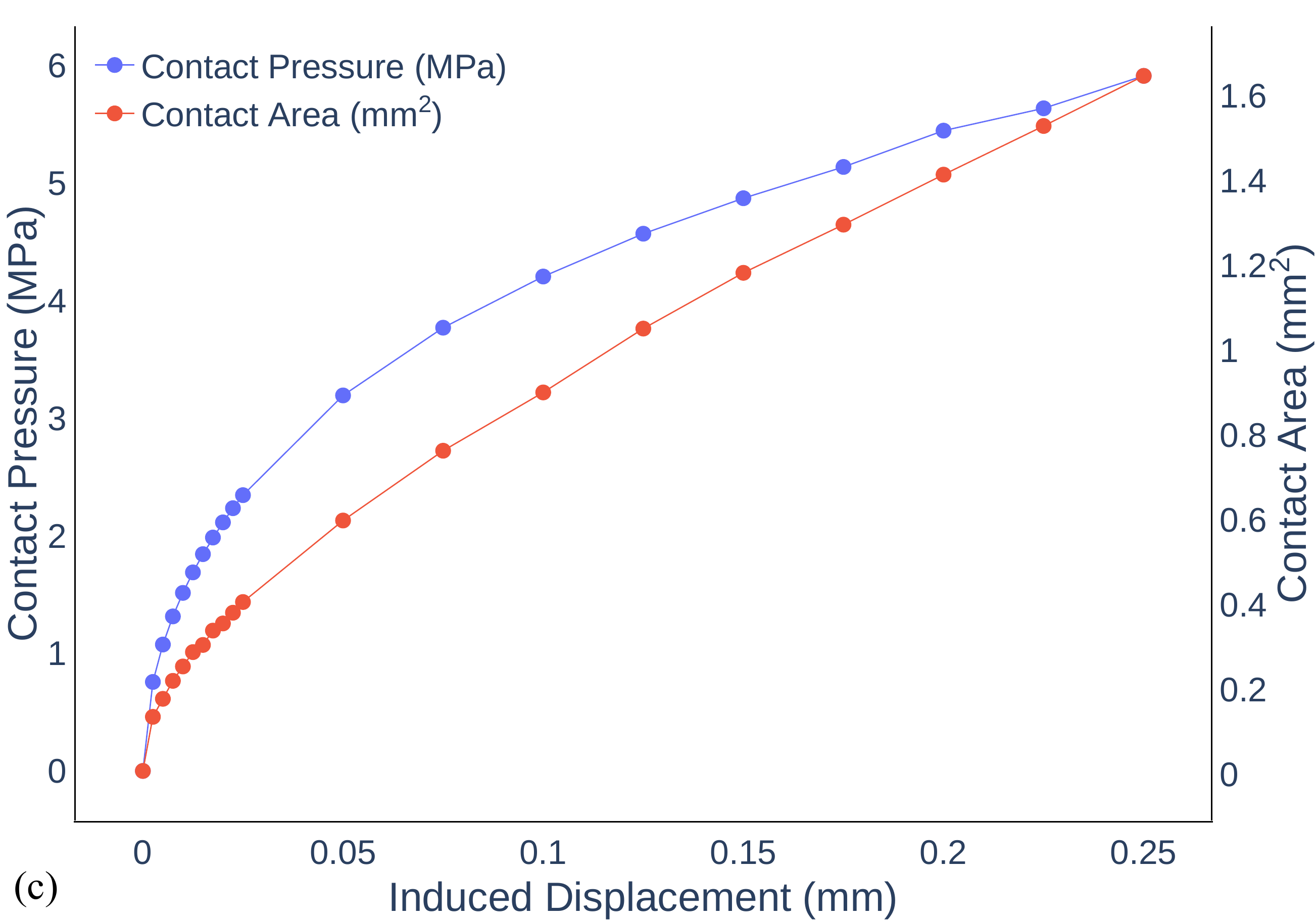}
\centering
\end{subfigure}

\caption{Case Study in Passive Adaptation to Loading: (a) Boundary Conditions (b) Final Deformed Shape (c) Contact Response}
\label{fig:BeamCaseStudy}
\end{figure}

It is our hypothesis that the use of auxetic metamaterials may provide functional advantages over traditional materials in the design of mechanisms whose working principle depends on passive adaptation to shape and loading. Here passive adaptation is defined as structural response of a system to external loading which improves design performance against a range of conditions without requiring active control. A mechanism with NPR will expand when tensile stresses are developed as a result of loading. If the mechanism involves compression against an external body due to contact, then the tendency of the auxetic structure to expand against that body will create a conformal interface whereby the mechanism deforms to the contour of the body. In principle, the expected result is an increase in contact area \cite{SoftAuxeticGripper}. Moreover, the expansion of the auxetic mechanism against the payload would result in the development of locally compressive stresses near the contact region, which is expected to increase reaction forces at the contact interface. Therefore, the two expected criteria that can be used to evaluate whether an auxetic structure will provide an advantage in adaptation to shape and loading are: (1) the development of tensile stresses and (2) contact against an external body which works to develop locally compressive stresses in the mechanism. To test this hypothesis we chose a simple robotic gripper as a case study. A simple case study of an auxetic sheet against an indenting body is shown in Figure~\ref{fig:BeamCaseStudy}, indicating an upward trend in both contacting area and contact pressure as the auxetic sheet is stretched.

Figure \ref{fig:Gripper_Cell}c illustrates the auxetic effect as it relates to the gripper case study. The force $F_{\textrm{in}}$ applied on the top surface of the gripper causes the jaws to curve inwards thereby exerting horizontal reaction force $F_{x}$ against the payload. As the jaws are loaded, the force in the vertical direction($F_y$) causes elongation of the gripper jaw in the y-direction. Since the jaws are auxetic, this vertical elongation also results in a horizontal expansion which results in the increase in the grip force $F_x$. This self-reinforcement may lead to the cellular auxetic gripper having a larger payload capacity when compared to non-auxetic designs. 
This paper therefore presents two primary objectives:
\begin{enumerate}
    \item Create an open-source computational framework for automated design of structures comprising parametric auxetic cells, by integrating cell parametrization, cellular system synthesis, FEA and surrogate-based optimization.
    \item Use this framework to explore the hypothesis of achieving a larger (gripping) reaction force on the loading with a auxetic cell-based structure compared to a structure given by topology optimization (with equivalent mass fraction).
\end{enumerate}

The remainder of this paper is structured as follows: Section \ref{sec:framework} describes the proposed auxetic metamaterial design framework and pertinent details of the framework components. Section \ref{sec:problem_def} details the robotic gripper case study and the corresponding design optimization formulation. Section \ref{sec:Results} contains the results from the case study. Concluding remarks are summarized in Section \ref{sec:Conclusion}.

\begin{figure*}[t]
\includegraphics[width= \linewidth]{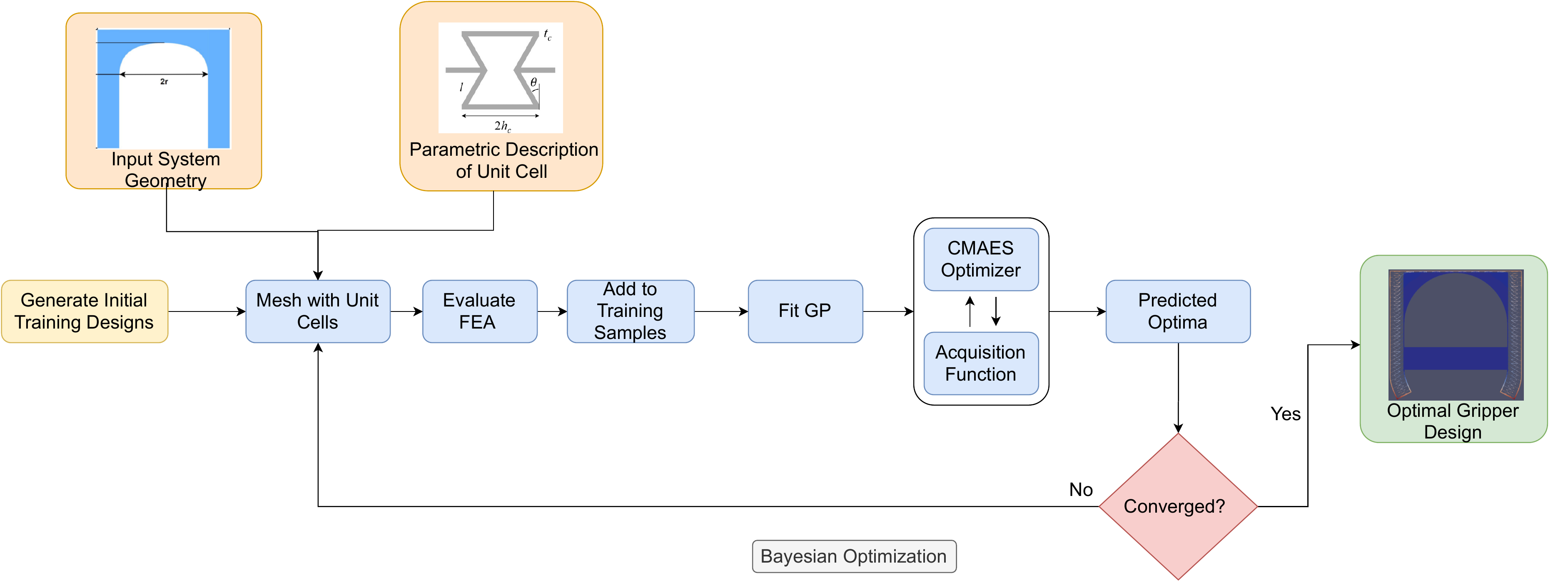}
\centering
\caption{Auxetic gripper design framework}
\label{fig:DesignFramework}
\end{figure*}

\begin{figure*}[]
\begin{minipage}{0.48\linewidth}
\centering
\captionof{table}{Design variables} \label{tab:design_variables}
    \begin{tabularx}{\linewidth}{>{\centering}X  >{\centering}X}
        \toprule
        Variable & Range \tabularnewline
        \midrule
        $e_{\textrm{offset}}$   & 0.1 mm to 1.0 mm \tabularnewline
        
        $h_c$  & 5.0 mm to 20.0 mm \tabularnewline 
        
        $\ell$  & 5.0 mm to 20.0 mm \tabularnewline
        
        $t_c$  & 0.2 mm to 1.0 mm \tabularnewline
        
        $\theta$ & $-60^{\circ}$ to $60^{\circ}$ \tabularnewline
    \bottomrule
    \end{tabularx}
\end{minipage} \hfill
\begin{minipage}{0.48\linewidth}
\centering
\captionof{table}{Fixed Design Parameters and Boundary Conditions} \label{tab:design_constants}
\begin{tabularx}{\linewidth}{>{\centering}X >{\centering}X}
\toprule
Parameter & Value \tabularnewline
\midrule
$\nu$ &  0.36 \tabularnewline

$E$  & 3750 MPa \tabularnewline

$d$ & 5.0 mm \tabularnewline 

$(m\times n)_{\textrm{top}}$ &  $2 \times 10$ \tabularnewline

$(m\times n)_{\textrm{sides}}$ &  $30 \times 2$ \tabularnewline

$v_{\textrm{applied}}$ &  -0.25 mm \tabularnewline

Payload Weight & 20 N \tabularnewline

$\theta_{\textrm{fixed}}$ &  $180^\circ$ \tabularnewline
\bottomrule
\end{tabularx}
\end{minipage}
\end{figure*}

\vspace{-0.3cm}




\section{Framework}
\label{sec:framework}

The proposed framework, shown in Figure~\ref{fig:DesignFramework} for the conceptual design of auxetic systems involves several steps. This framework focuses on auxetic systems using two-dimensional re-entrant honeycomb unit cells, but can be easily adapted in other auxetic unit cell configurations. Firstly, a structured meshing algorithm is employed to convert the initial geometry into a finite system of the initial auxetic unit cells. The meshing algorithm is designed to fill the interior of the input design with a specified number of unit cells (set as a \textit{hyperparameter}) and leaves a strip of solid material along the boundaries. A multi-fidelity Bayesian optimization is then performed to find the unit cell parameters which yield the optimal design. The use of Bayesian optimization decreases the computational cost of the optimization and allows for this framework to be applied to problems with a very expensive objective function. 

\subsection{Structured meshing of gripper geometry with auxetic unit cells}

One of the challenges associated with using auxetic metamaterials in the design of mechanisms is that manually incorporating and repeating unit cells in the structure is time consuming and tedious. As such, an automated methodology is proposed to mesh the geometry with auxetic unit cells. Two-dimensional honeycomb cells have a geometry that is well defined and can be bounded by a rectangle for all possible shapes. Therefore, the most natural automated framework for meshing a structure with unit cells is with a structured $m \times n$ rectangular mesh. 


There are several approaches which may be used to generate a structure with a valid mesh of auxetic cells. The algorithm begins by using a rectangular region of specified size within the mechanism as an integer value of unit cells, that is an $m \times n$ finite system of cells separated from the boundary by at least some edge offset, which itself can either be fixed or made a design variable. If the unit cell design and outer dimensions of the mechanism are to be fixed, then the allowable $m \times n$ can easily be calculated. If the unit cell design is to be varied and the outer dimensions are bounded, then constraints can be placed on the final design optimization to limit unit cell design parameters which would result in an infeasible design. If the unit cell design is to be varied and the allowable outer dimensions are relatively unconstrained and can be varied, then the design optimization can solve with no direct constraints on mechanism dimensions, with outer dimensions being constrained only implicitly by the range of unit cell design variables being considered. Once regions inside the mechanism are defined using one of these three approaches, the remainder of the mechanism can be filled in with solid material. For simple cases this can be accomplished by applying an edge offset from the vertices of the rectangular finite system(s). For more complex cases a series of Boolean operations may need to be performed on the CAD geometry to fill the regions between the finite systems and boundaries of the mechanism with solid material.

\subsection{Finite element analysis (FEA)}

High-fidelity function evaluations used in the  optimization problem are built within the open-source finite element analysis code FEniCS for seamless integration into the design framework \cite{LoggWells2010,LoggEtal_10_2012,AlnaesEtal2014,alnaes2010}. The finite element model (FEM) is a two-dimensional plane stress analysis and geometry is created and meshed with first order triangular elements using Gmsh \cite{Gmsh}. The FEM was developed in a way which assumes the material of the auxetic mechanism and payload are the same, but in general this is not necessary. Both the auxetic mechanism and payload are explicitly modeled in the FEM to capture boundary stiffness. The FEM is developed for the solution of small deformation linear elastic problems. Plasticity can be accounted for through the appropriate selection and implementation of a nonlinear constitutive model, but the inability to model geometric nonlinearity is a limitation of the FEM used in this analysis.

Since the scope of this paper is limited to finding the optimal unit cell parameters and not change the outer geometry, a conformal mesh is implemented between the payload and the mechanism which is representative of a bonded contact with constant area through loading. For the case of a simple rectangular bar acting as the payload, the stiffness can be easily calculated analytically as $A_pE/L_p$ and reaction forces from actuation of the auxetic mechanism at the conformal interface can be calculated with knowledge of the payload deformation. Frictional, frictionless, or contact formulations which otherwise allow for separation are nonlinear in nature and have not been incorporated into the FEniCS code. For the study of passive adaptation to shape, more refined contact modeling which allows for separation would be prudent.

The FEM is also used to evaluate whether a trial design satisfies the constraints of the optimization problem. Imposed constraints can include limitations on maximum stress in the assembly. Unrealistically stiff fixed boundary conditions can cause artificially high values of stress near these boundaries. In addition, stress concentrations will be observed in the re-entrant honeycomb unit cell if edge fillets (curved surface of a constant radius that joins to surfaces to reduce stress at corners) are not included in the FEM. Care should be taken when imposing constraints on stress to account for artificial stresses resulting from idealizations in the FEM.

\subsection{Design optimization}

Here, we make use of a Bayesian optimization algorithm since it uses an in-built Gaussian process surrogate model which greatly reduces the computational cost of the optimization \cite{greenhill2020bayesian}. We use the constrained expected improvement \cite{gardner2014bayesian} as the acquisition function and a CMAES algorithm \cite{hansen2016cma} to optimize the acquisition function.

Using a  global optimizer is advantageous since the use of local optimizers may result in the algorithm getting stuck in local minima and may give results which are marginal when compared with traditional materials due to the nonlinearity of the design problem. Given that mesh sizes in auxetic mechanisms usually are very large, surrogate modeling reduces the computational cost associated with evaluating the FEM. Table \ref{tab:Bayesopt} contains the optimization settings used.


\begin{table}[]
    \caption{Bayesian optimization settings}
    \label{tab:Bayesopt}
    \centering
    \begin{tabularx}{\linewidth}{>{\centering}X >{\centering}X}
    \toprule
        Parameter & Value \tabularnewline
        \midrule
       Acquisition Function  &  Constrained Expected Improvement \tabularnewline
       Initial Number of Training Samples  &  500 \tabularnewline
       New Samples per Iteration & 1 \tabularnewline
       Number of Iterations & 100 \tabularnewline
       Acquisition Function Optimizer  & CMAES \tabularnewline
       CMAES Population & 80 \tabularnewline
       \bottomrule
       
    \end{tabularx}

\end{table}


\begin{figure}[]
\centering
    \includegraphics[width=0.42\linewidth]{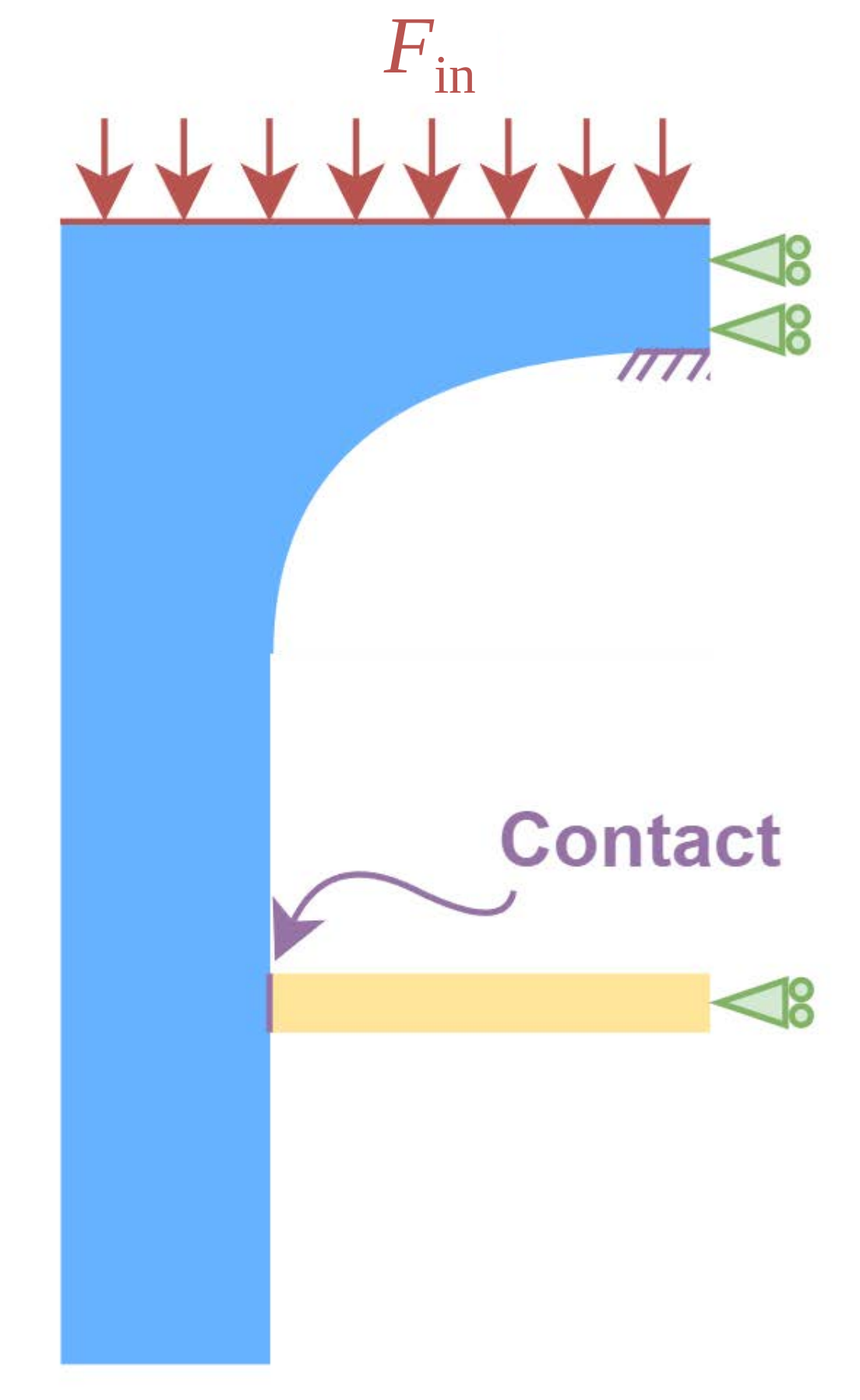}
\centering
\captionof{figure}{Validation FEM boundary conditions}
\label{fig:BCs_validation}
\end{figure}

\begin{table}
    \captionof{table}{Validation FEM settings}
    \label{tab:validation_fea}
    \centering
    \begin{tabularx}{\linewidth}{>{\centering}X >{\centering}X}
    \toprule
      Analysis Setting & Value \tabularnewline
      \midrule
      Type  &  Static, Plane Stress \tabularnewline
      Solver Controls & Large Deflection \tabularnewline
      Element Type & Quadrilateral 8 Dominant \tabularnewline
      Contact Type & Bonded \tabularnewline
      Input Load ($F_{\textrm{in}}$) & 50 N \tabularnewline
      Max Payload Weight & 20 N \tabularnewline
      \bottomrule
    \end{tabularx}
\end{table}

\section{Robotic Gripper Case Study Formulation}
\label{sec:problem_def}
The proposed optimal design framework has been tested using a soft robotic gripper case study. Our robotic gripper design is also compared to a gripper designed using topology optimization of the same initial design envelope to assess advantages and disadvantages of the proposed framework against a well established and very general optimal design technique.

The gripper design begins with an initial envelope defined by the dependent variables shown in Figure~\ref{fig:Gripper_Cell}(a) with assumed symmetry about the vertical centerline. The design is dependent on the two-dimensional re-entrant honeycomb cells and finite system size with unit cell design variables parameterized according to Figure~\ref{fig:Gripper_Cell}(b).

\begin{figure}[h!]
\includegraphics[width=0.75\linewidth]{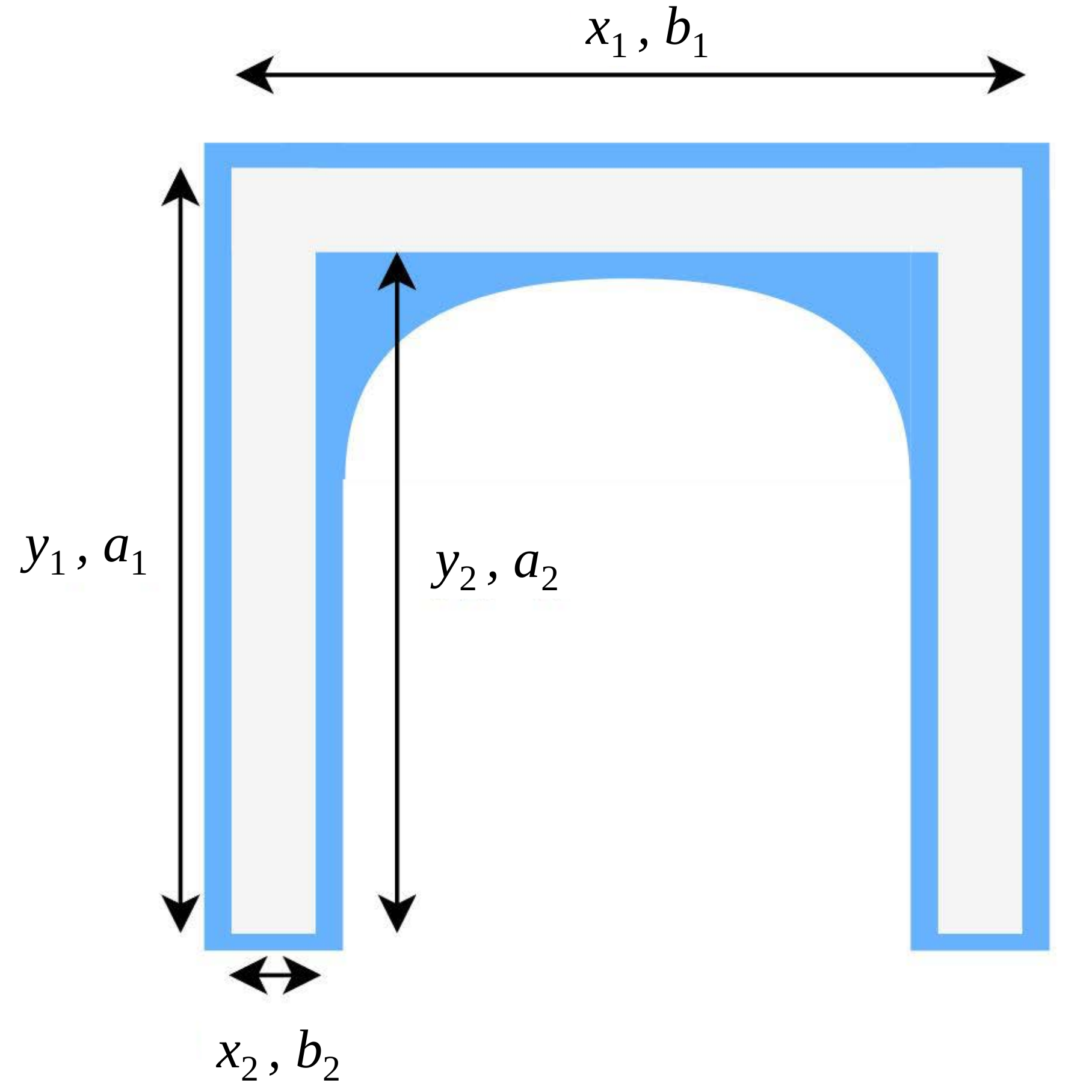}
\centering
\caption{Gripper mesh parameters}
\label{fig:GripperMesh}
\end{figure}

For the gripper case study, it was assumed that the outer dimensions are relatively unconstrained and that they could be defined simply by creating three unit cell grids of size $m_i \times n_i$ on the left, top, and right side of the gripper geometry. The outer dimensions of the gripper are defined by an edge offset which itself is a design variable. Outer dimensions are limited only by the minimum and maximum bounds on $m,n$, the unit cell design $h,\ell,\theta$, and the edge offset. The unit cell finite systems are also assumed symmetric about the vertical centerline. Inner dimensions of the gripper filled with unit cells are defined by $x_1,x_2,y_1,y_2$ as shown in Figure~\ref{fig:GripperMesh}. These values are lengths calculated based on knowledge of unit cell design and edge offset.

Once the gripper geometry has been meshed with honeycomb cells, the framework proceeds to solving the design optimization problem, which is mathematically expressed in Equations \ref{eq:design_obj} through \ref{eq:design_con3}
\begin{equation} \textrm{min}_x \ \ f(\vec{x}): -\frac{u_{\textrm{out}}}{v_{\textrm{out}}} \label{eq:design_obj} \end{equation}
subject to the constraint
\begin{equation} g_1(\vec{x}): \left[(m_2^x-\delta) - (m_1^x + \delta) - t_c/2 \right] > 0. \label{eq:design_con3} \end{equation}
where
\begin{subequations}
  \begin{gather}
    m_1^x = -h+l \sin{(\theta)}+ \delta/2\\
    m_2^x = h-l \sin{(\theta)}-\delta/2\\
    \delta = t_c/\cos{\theta}\\
    \overrightarrow{x} = \left[e_{\text{offset}}, h_c, l, t_c, \theta \right]
  \end{gather}
\end{subequations}
Where, $u_{\textrm{out}}$ and $v_{\textrm{out}}$ are the mean of the absolute x and y displacements of the jaws respectively. 
The constraint $g_1$ is imposed to prevent overlap between the honeycomb cell diagonal ribs at the vertical centerline with $t_c/2$ representing the minimum required gap between the diagonal ribs. Constraints on the design variables are summarized in Table \ref{tab:design_variables} and constant parameters used in the design are summarized in Table \ref{tab:design_constants}. The payload length is equal to the perpendicular distance between the vertical gripper jaws. Both the auxetic gripper and the gripper designed using topology optimization use PLLA II as the material. The payload is modeled as a rectangular bar between the gripper jaws with equivalent stiffness $A_pE/L_p = w_p d E/L_p$ and interacts with the gripper through a conformal mesh which is representative of a constant area bonded contact. For this case study, it is assumed that the payload width $w_p = h_g/3$ is proportional to the gripper dimensions and is centrally located in the gripper jaws.

\begin{figure}[h!]
\includegraphics[width=0.7\linewidth]{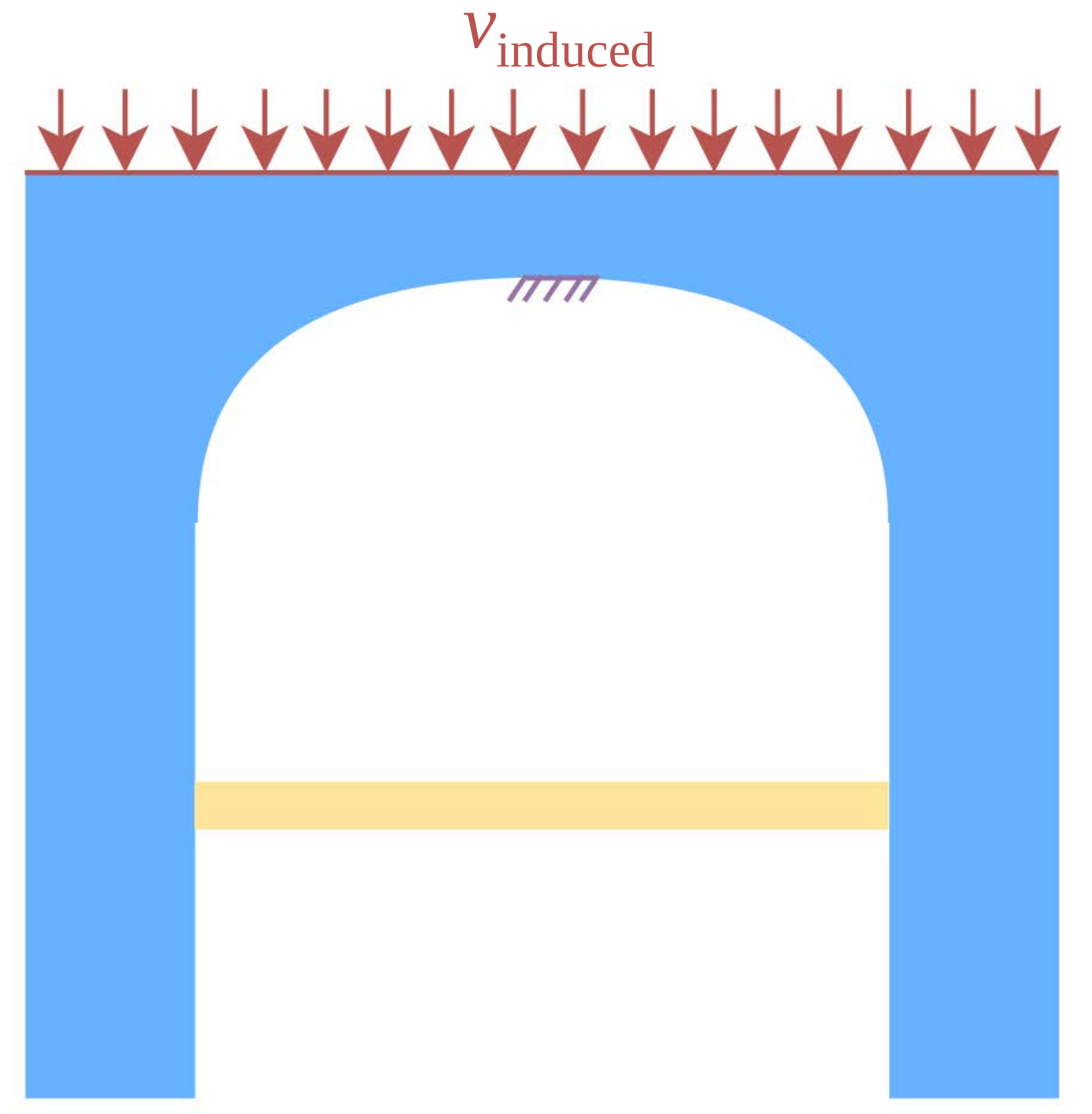}
\centering
\caption{Gripper FEM boundary conditions}
\label{fig:BCs}
\end{figure}

Throughout the optimization, both the solid gripper envelope and the unit cell variables are changed to achieve an optimal design. The objective function and constraints are approximated using a surrogate model to reduce the computational cost of the optimization, and the final design is verified by evaluating the objective function with a high fidelity FEM. Symmetry constraints are imposed on the vertical centerline of the gripper to further reduce the cost of evaluating the objective function for the topology optimization baseline and validation cases only. The symmetry constraints are applied to both the gripper and the payload. The boundary conditions for the FEM are pictorially shown in Figure~\ref{fig:BCs}. There are four separate boundary conditions used in the FEM. First, symmetry about the vertical centerline of the gripper is imposed by using constraints against normal displacement, $u=0$, for the topology optimization and validation analyses only. Second, displacement is fixed on the inner edge of the gripper at the elliptical cutout, $u=v=0$. Third, a uniform displacement is applied to the top edge of the gripper which is representative of the actuation input. Finally, constant forces are imposed at the center of the payload representative of its weight. For the topology optimization reference design, all enveloping edges of the gripper are treated as exclusion zones, as is the payload body.

The optimal design was generated using the proposed framework and by setting the design shown in Figure~\ref{fig:Gripper_Cell}(a) as the initial geometry. This design is also compared to two designs generated using topology optimization one with a volume fraction of 8\% - similar to the volume fraction of the optimal auxetic design and a second design with 25\% volume fraction. The settings used for topology optimization are listed in Table \ref{tab:TopOpSettings}. A validation FEM with parameters shown in Table \ref{tab:validation_fea} and boundary conditions as shown in Figure \ref{fig:BCs_validation} is then performed on all these designs. The resulting grip forces, as well as the peak stress values are tabulated in Table \ref{tab:results}.

\begin{figure*}
    \centering
    \includegraphics[width=\linewidth]{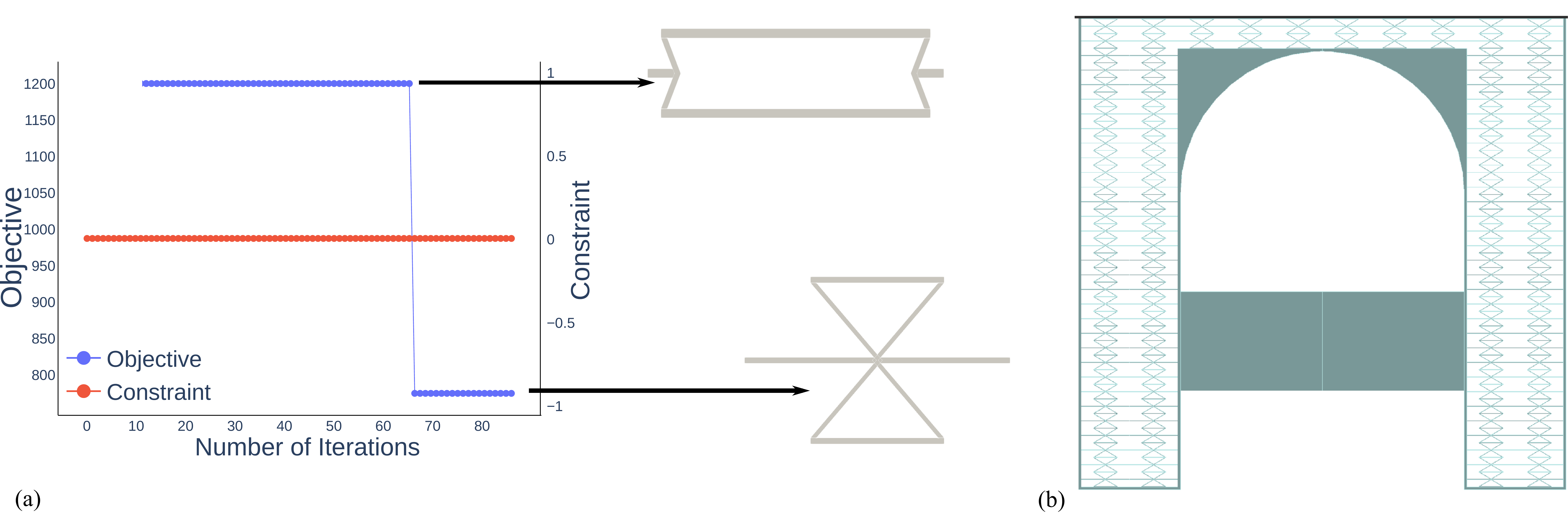}
    \caption{Optimization results: (a) Convergence history, and (b) Optimal gripper geometry}
    \label{fig:conv_hist}
\end{figure*}

\begin{figure*}[t]
\includegraphics[width=\linewidth]{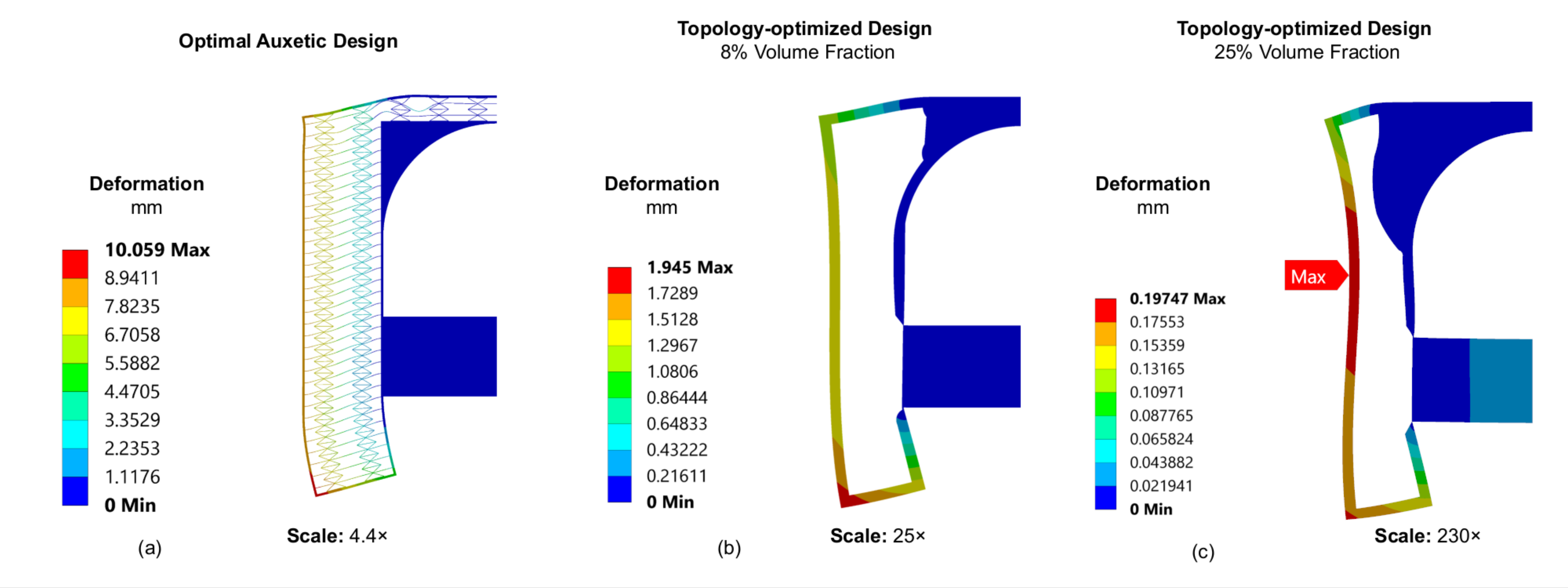}
\centering
\caption{Deformation contours in mm corresponding to: (a) Cellular auxetic design, (b) Topology-optimized design with 8\% volume fraction (i.e., with mass equivalent to cellular auxetic design), and (c) Topology-optimized design with 25\% volume fraction}
\label{fig:deformation_plots}
\end{figure*}

\begin{table}[h!]
    \caption{Topology optimization settings}
    \label{tab:TopOpSettings}
    \centering
    \begin{tabularx}{\linewidth}{>{\centering}X >{\centering}X}
    \toprule
    Setting & Value \tabularnewline
    \midrule
    Objective (minimize) & Compliance  \tabularnewline
    Max Number Iterations & 100 \tabularnewline
    Convergence Accuracy & 0.1\% \tabularnewline
    Required Number Iterations & 27 \tabularnewline
    Retain Threshold & 0.1 \tabularnewline
    \bottomrule 
    \end{tabularx}
\end{table}

\section{Results and Discussion}
\label{sec:Results}
Figure \ref{fig:conv_hist} shows the convergence history of the optimization algorithm, along with the auxetic unit cell designs at different points during the solution history. Due to the time constraints associated with the initial draft, the optimization was stopped at 80 iterations. We expect further iterations of the optimization to produce better auxetic designs. The final optimized gripper geometry along with the unit cells is shown in Figure~\ref{fig:conv_hist}.

From the results it can be observed that the auxetic design has a higher grip force than both the topology optimized designs. Of the three designs the 8\% volume fraction design performs the poorest in terms of grip force. The 25\% volume fraction design has a larger grip force than the 8\% volume fraction design, but is less than the auxetic design. 

From the deformation plots in Figure~\ref{fig:deformation_plots}, we can see that the topology optimized designs tend to have a thinner ``jaws'' and less material between the top surface and the gripper jaws. This behavior may change if additional constraints are imposed during the topology optimization process. The topology optimization results also needed additional manual post-processing on the outer contour to perform the validation FEM. The auxetic design on the other hand does not require manual post-processing is also geometrically simple enough to fabricate via 3D printing. A disadvantage to the re-entrant auxetic design is the peak stress in greater than the topology optimized designs. This could easily be addressed by using a smoothed re-entrant unit cell which limits stress concentrations.


\begin{figure*}[]
\begin{subfigure}{0.6\linewidth}
\includegraphics[width= \linewidth]{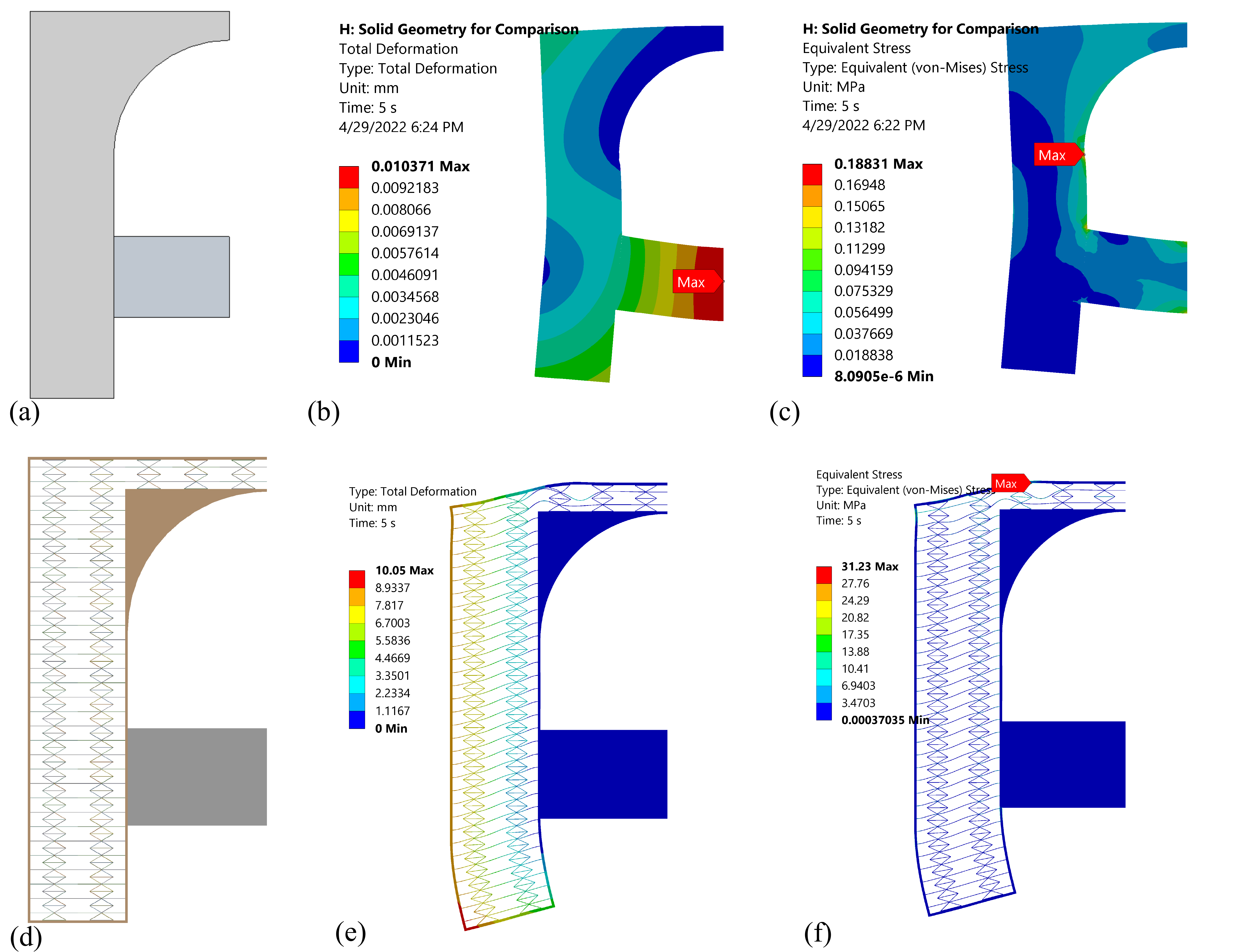}
\centering
\end{subfigure} \hfill
\begin{subfigure}{0.39\linewidth}
\includegraphics[width= \linewidth]{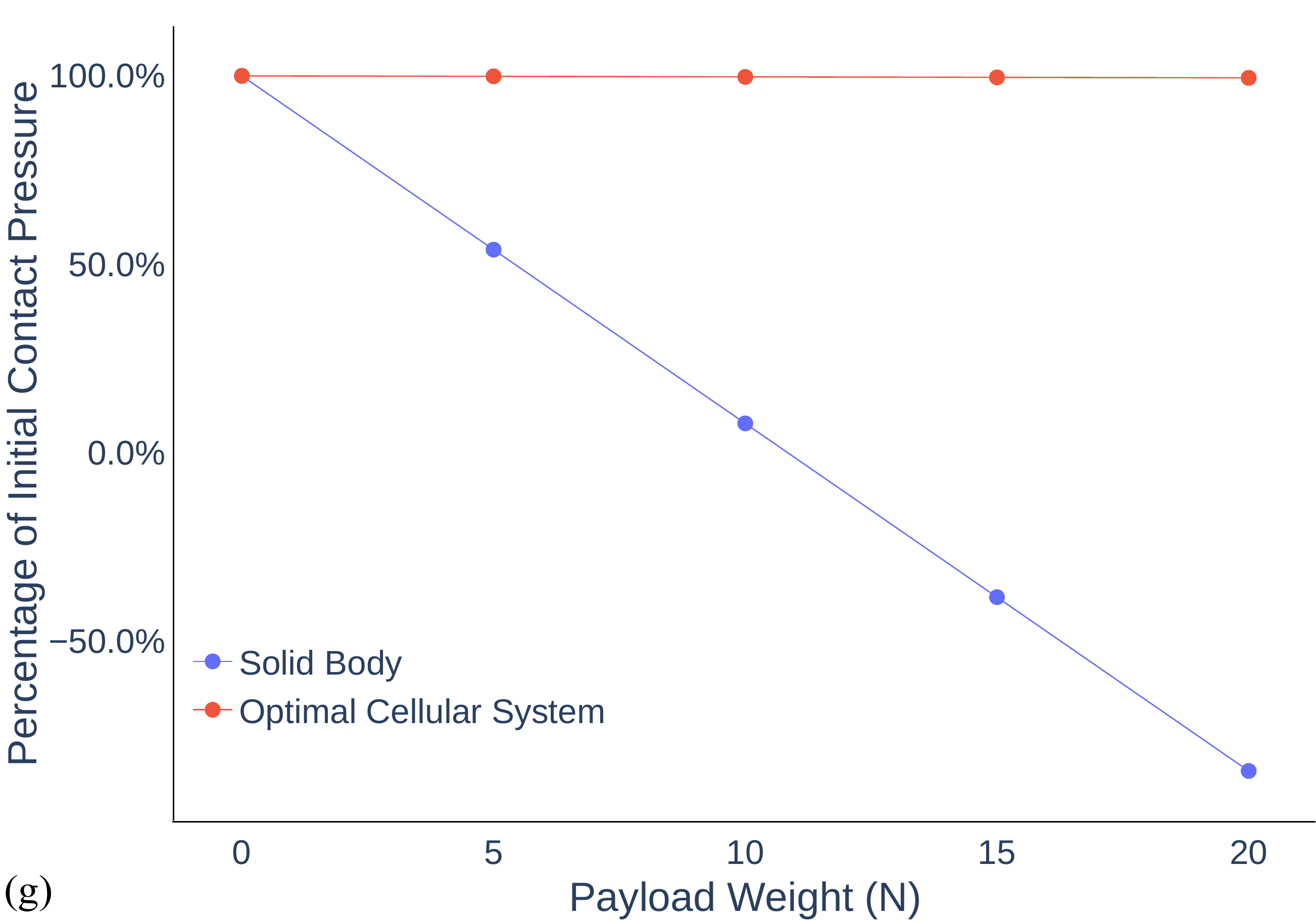}
\centering
\end{subfigure}

\caption{Passive Adaptation to Loading: (a) Solid Geometry (b) Total Deformation Plot of Solid Geometry at Payload 20N (c) Equivalent Stress Plot of Solid Geometry at Payload 20N (d) Optimal Cellular Geometry (e) Total Deformation Plot of Optimal Cellular Geometry at Payload 20N (f) Equivalent Stress Plot of Optimal Cellular Geometry at Payload 20N (g) Change in Percent Contact Pressure with Change in Payload Weight}
\label{fig:PassiveAdaptCompare}
\end{figure*}

\begin{table}[]
    \caption{Comparing auxetic-cell based system with topology optimization based system}
    \label{tab:results}
    \begin{threeparttable}
    \centering
    \begin{tabularx}{\linewidth}{>{\centering}X >{\centering}X >{\centering}X >{\centering}X}
    \toprule
    Quantity & Cellular Auxetic Design (8\%) & Topology Optimization Design (8\%) \tnote{1} & Topology Optimization Design (25\%) \tabularnewline
    \midrule
    Reaction Force \tnote{2} & 1.65 N  & 0.41 N & 0.85 N\tabularnewline
    Peak von-Mises Stress \tnote{3} & 31.17 MPa & 7.41 MPa & 2.01 MPa \tabularnewline
    \bottomrule 
    \end{tabularx}
    \begin{tablenotes}
    \item[1] Optimal auxetic design and topology optimization design with 8\% volume fraction have equivalent mass
     \item[2] Higher value is better
     \item[3] Lower value is better
   \end{tablenotes}
       \end{threeparttable}
\end{table}

\begin{table}[]
    \caption{X and Y Displacement of Gripper Jaws}
    \label{tab:jawdisp}
    \centering
    \begin{tabularx}{\linewidth}{>{\centering}X >{\centering}X >{\centering}X >{\centering}s >{\centering}s }
    \toprule
    Gripper Type & X Disp. & Y Disp. & Input Force & Payload Weight \tabularnewline
    \midrule
    Topology Optimization - 25\% & 1.4 $\times 10^{-4}$ & 1.7 $\times 10^{-5}$ & 50 N & 20 N  \tabularnewline
    Topology Optimization - 8\% & 1.5 $\times 10^{-3}$ & 8.3 $\times 10^{-6}$ & 50 N & 20 N  \tabularnewline
    Optimal Cellular Design & 5.2 $\times 10^{-3}$ & 1.9 $\times 10^{-4}$  & 50 N & 20 N  \tabularnewline
    \bottomrule 
    \end{tabularx}
\end{table}

The design obtained from the optimization framework is compared with a solid gripper of the same outer dimensions and under the same loading conditions for evaluation of passive adaptation to loading. Table \ref{tab:jawdisp} shows the X and Y displacement of the jaws of both the topology optimized designs and the optimal cellular design. 
Figure ~\ref{fig:PassiveAdaptCompare}(a-f) graphically shows the geometry (design), deformation, and stresses under 50 N of actuation loading on the top surface and 10 N of payload applied at the centerline (symmetry plane). The loading is applied over several steps, with a ramp increase in the actuation force over the first load step followed by incremental increases in payload. This is similar to how loading of an actual robotic gripper would occur, where the gripper actuation would cause normal gripping forces to develop prior to shear (frictional) loading to carry the payload. The finite element models were used to quantify whether the auxetic system exhibited passive adaptation to loading, which can be characterized by the slope of the gripping force to payload weight line, as shown in Figure~\ref{fig:PassiveAdaptCompare}(g). The finite element model for evaluation of passive adaptation to loading used bonded contact, as indicated in Table~\ref{tab:validation_fea}, to evaluate passive adaptation to loading. 




It can be seen from Figure~\ref{fig:PassiveAdaptCompare}(g) that the cellular auxetic gripper has a negligible slope (-0.00263 N/N) compared to the solid gripper's slope (-0.4515 N/N), indicating that the auxetic design does not suffer from a significant reduction in normal gripping force against the contacting body due to downward loading of the payload, unlike the solid gripper. The decrease in gripping force is approximately $172\times$ greater for the solid gripper than for the auxetic gripper. At approximately 5.6 N of payload, the solid gripper x-direction gripping force switches signs to become negative. Since bonded contact is assumed in the FEM, this is physically indicative of separation between the payload and gripper jaws, where frictional forces are no longer developed and no load can be carried by the gripper. Therefore the auxetic system is said to exhibit \textit{passive adaptation to variable loading} relative to the solid gripper. The auxetic passive adaptation can be attributed to the fundamentally different deformed shape of the cellular and solid designs. That is, the jaws of the cellular gripper bend inwards, whereas for the solid design the jaws bend outward.

\section{Concluding Remarks}
\label{sec:Conclusion}
In this paper, we explore the feasibility of using auxetic cell-based metamaterial to build passively load-adaptive structures, and developed a new end-to-end design automation framework for this purpose. The presented framework is demonstrated by a design implementation of an auxetic robot-gripper with a fixed force input. Therein, first the inner geometry of the gripper was discretized and filled with unit cells. Following which, the parameters of the unit cell were optimized with an objective function designed to maximize the horizontal displacement at the jaws while simultaneously minimizing the vertical displacement. To this end, an efficient Bayesian optimization algorithm with constrained expected improvement as the acquisition function and CMAES as the optimizer was utilized. State-of-the-art open-source tools were used to create this entire synthesis-analysis-optimization framework in order to allow ready adoption and usage by the community. 

The optimization outcome is validated by computing the reaction forces and stresses using a higher fidelity commercial FEM software (ANSYS). Standard topology optimization was then carried out twice in this software for comparison, once with the volume fraction equal to that of the optimal auxetic design (which was 8\%) and another with a higher, commonly used volume fraction of 25\%. The validation FEM was also run on the aforementioned designs and the results demonstrated that the auxetic gripper provides a higher reaction force when compared to both of the topology optimization results, while also not requiring any manual post-processing (which is attributed to the use of the parametric auxetic unit cells). Future work will include generalization of the framework to explore the degree of auxetic characteristics exhibited at the unit cell level when multiple trade-off functionalities such as space filling, high reaction force, higher contact area, lower mass and smaller median stresses within are desired. Lastly, extending the implementation to more realistic gripper and robotic component shapes will further elucidate the advantages and limits of such parametric cell-based metametarial structures. 


\bibliographystyle{unsrt}
\bibliography{asme2e}

\end{document}